\documentclass[twocolumn]{aastex62}
\usepackage{graphics,epsf}
\usepackage{amsmath}                % American Mathematical Society package
\usepackage{amsfonts}               % American Mathematical Society fonts
\usepackage{amssymb}                % American Mathematical Society symbol
\usepackage{epsfig}                 % EPS figures
\usepackage{appendix}
\usepackage{graphicx}
\usepackage{float}
\usepackage{color}
\usepackage{multirow}
\usepackage{colortbl}
\usepackage[para,online,flushleft]{threeparttable}

\newcommand{\cm}{{~\rm cm}}

\newcommand{\km}{{~\rm km}}
\newcommand{\s}{{~\rm s}}

\newcommand{\g}{{~\rm g}}

\newcommand{\erg}{{~\rm erg}}
\newcommand{\yr}{{~\rm yr}}

\newcommand{\Mpc}{{~\rm Mpc}}
\newcommand{\eV}{{~\rm eV}}

\newcommand{\AU}{{~\rm AU}}

\begin{document}

\title{Double common envelope jets supernovae (CEJSNe) by triple-star systems}

%\correspondingauthor{Ealeal Bear, Noam Soker}
\email{soker@physics.technion.ac.il}

\author{Noam Soker}
\affiliation{Department of Physics, Technion – Israel Institute of Technology, Haifa 3200003, Israel}
\affiliation{Guangdong Technion Israel Institute of Technology, Guangdong Province, Shantou 515069, China}

\begin{abstract}
I propose a new type of common envelope jets supernova (CEJSN) events where instead of a single neutron star (NS; or a black hole; BH) a tight binary system of a NS and a main sequence star enters a common envelope evolution (CEE) with a red supergiant. 
The NS and the main sequence star of the tight binary system merge inside the red supergiant envelope and enter a CEE of their own. The NS accretes some mass through an accretion disk and launches jets that explodes the main sequence star. I estimate that the two jets that the NS launches at this phase carry an energy of $\approx 10^{52} \erg$, about the same order of magnitude as the energy that 
 the jets will carry when the NS or its BH remnant will enter the core in a later phase. For that, I term the entire event a double CEJSN. The outcome of the double CEJSN is a very long, months to years, and very energetic event, a total energy of $\approx 10^{52} - 10^{53} \erg$, that will be observationally classified as a peculiar super-energetic event. I crudely estimate that new transient surveys should detect about one CEJSN event from a triple star system per year.  
\end{abstract}


% ==========================================================
\section{Introduction} 
\label{sec:intro}
% ==========================================================

In a common envelope jets supernova (CEJSN) event a neutron star (NS) or a black hole (BH) accretes mass of a red supergiant (RSG) star   from within its envelope and launches jets as it spirals-in in a common envelope evolution (CEE; e.g., \citealt{Gilkisetal2019, SokeretalGG2019, GrichenerSoker2019a, Schroderetal2020, GrichenerSoker2021}). The non-uniform density in the RSG envelope implies sufficiently large specific angular momentum of the mass that the NS/BH accretes to form an accretion disk around the NS/BH (e.g.,  \citealt{ArmitageLivio2000, Papishetal2015, SokerGilkis2018}; for recent hydrodynamical studies of disk formation see, e.g., \citealt{LopezCamaraetal2019, LopezCamaraetal2020MN}, and for a discussion on the formation of the accretion disk with more references see \citealt{Hilleletal2021}).  The density gradient in the envelope causes the NS/BH to accrete more mass from the denser side. The accreted gas from the denser side and the accreted gas from the other side of the NS/BH have opposite signs of the angular momentum around the NS/BH. Since more mass is accreted from the denser side, the angular momenta of the two sides do not cancel each other and the accreted mass has a net angular momentum. Because of the very small radius of the NS/BH the specific angular momentum of the accreted matter is sufficient to form an accretion disk around the NS/BH.  

The spiralling-in NS/BH ends in one of two possibilities. It might eject the entire envelope before it reaches the core, such that the dynamical friction ceases and the spiralling-in processes stops  (e.g., \citealt{SokeretalGG2019}).   At a later time the core explodes as a CCSN and the remnant is a system of two NSs, or NS + BH, or two BHs, bound or unbound  (e.g., \citealt{SokeretalGG2019}).   More relevant to the present study is the case where the NS/BH reaches the core and continues with the CEE inside the core. The last powering phase is when the NS/BH destroys the core and the destroyed core gas forms a massive accretion disk that launches energetic jets (e.g., \citealt{GrichenerSoker2019a}). The observed CEJSN event might be similar to a CCSN, but it is likely to be more energetic with explosions energy of up to $\simeq {\rm several} \times 10^{52} \erg$, might last for a longer time of up to few years, typically contains a more massive circumstellar matter (CSM) of up to several solar masses, and in general it would be a peculiar supernova (SN; e.g., \citealt{SokeretalGG2019, Schroderetal2020}).  The peculiarity might be in the long lasting and energetic light curve, and/or in several peaks in the light curve, and/or in abnormal composition due to nucleosynthesis inside the jets that the NS/BH launches while inside the core (e.g., \citealt{GrichenerSoker2019a}), and/or a highly non-spherical explosion (e.g., \citealt{SokeretalGG2019}). As well, the NS/BH merger with the core might be a source of gravitational waves (e.g., \citealt{Ginatetal2020}).  
\cite{SokerGilkis2018} attributed the enigmatic SN~iPTF14hls \citep{Arcavietal2017}, and now also the similar SN~2020faa \citep{Yangetal2021} to a CEJSN event. \cite{SokeretalGG2019} developed the  polar CEJSN scenario where the jets that the NS launches as it spirals-in inside the RSG envelope clear two opposite polar regions, allowing the later jets that the NS launches as it spirals-in inside the core to expand freely. They suggested this scenario to explain the 
fast-rising blue optical transient AT2018cow (e.g., \citealt{Prenticeetal2018, Marguttietal2019}). 
  
A second key process that allows the CEJSN, the first one being the formation of an accretion disk around the NS/BH, is the efficient neutrino-cooling of the accreted mass when the accretion rate onto a NS is $\dot M_{\rm acc} \ga 10^{-3} M_\odot \yr^{-1}$ \citep{HouckChevalier1991, Chevalier1993, Chevalier2012}. In the case of a BH the accreted gas can in addition carry a large fraction of the energy into the BH (e.g., \citealt{Pophametal1999}). Namely, for accreting NS/BHs in a CEE the Eddington accretion limit is not relevant and the NS/BH might accrete mass at a very high rate, i.e., up to a tenth of a solar mass per second.   
The high accretion rate implies that the accretion disk might launch  energetic jets that in turn host high-energy processes, such as r-process nucleosynthesis \citep{Papishetal2015, GrichenerSoker2019a, GrichenerSoker2019b} and the formation of high-energy ($\approx 10^{15} \eV$) neutrinos when the accretor is a BH \citep{GrichenerSoker2021}. 
As well, the jets that the NS/BH launches unbind large amounts of envelope mass and can supply extra energy source to make the CEE efficiency parameter to be $\alpha_{\rm CE} > 1$, as some scenarios require (e.g. \citealt{Fragosetal2019, Zevinetal2021, Garciaetal2021}).

The increasing number of sky surveys is observing more and more peculiar SNe (e.g., \citealt{GalYam2012, GalYam2019, Grahametal2019, Holoienetal2019}). 
The goal of the present study is to further explore evolutionary routes that might lead to peculiar super-energetic supernovae (SNe) that are not pair instability SNe. Being super-energetic, i.e., with explosion energy of $E_{\rm exp} > 3 \times 10^{51} \erg$, these cannot be driven by the commonly studied delayed neutrino mechanism. Rather, these SNe must be driven by jets. The large fraction of massive stars in triple (or more) star systems (e.g., \citealt{Sanaetal2014, MoeDiStefano2017}) motivate this study of CEJSNe in triple-star systems. One such evolutionary route, the double CEJSN, is the subject of the present study. 

I describe the double CEJSN scenario in section \ref{sec:Scenario}, and quantify some of its parameters in section \ref{sec:Powering}. In section \ref{sec:rate} I estimate the rate of double CEJSN event relative to CCSN events. I summarise in section \ref{sec:Summary}.

% ==========================================================
\section{The double CEJSN scenario} 
\label{sec:Scenario}
% ==========================================================

\begin{figure*}%[ht]
 %\centering
%  \vskip -3.00 cm
 %\hskip -1.00 cm
\includegraphics[trim=0.0cm 0.0cm 0.0cm 2.0cm ,clip, scale=0.80]{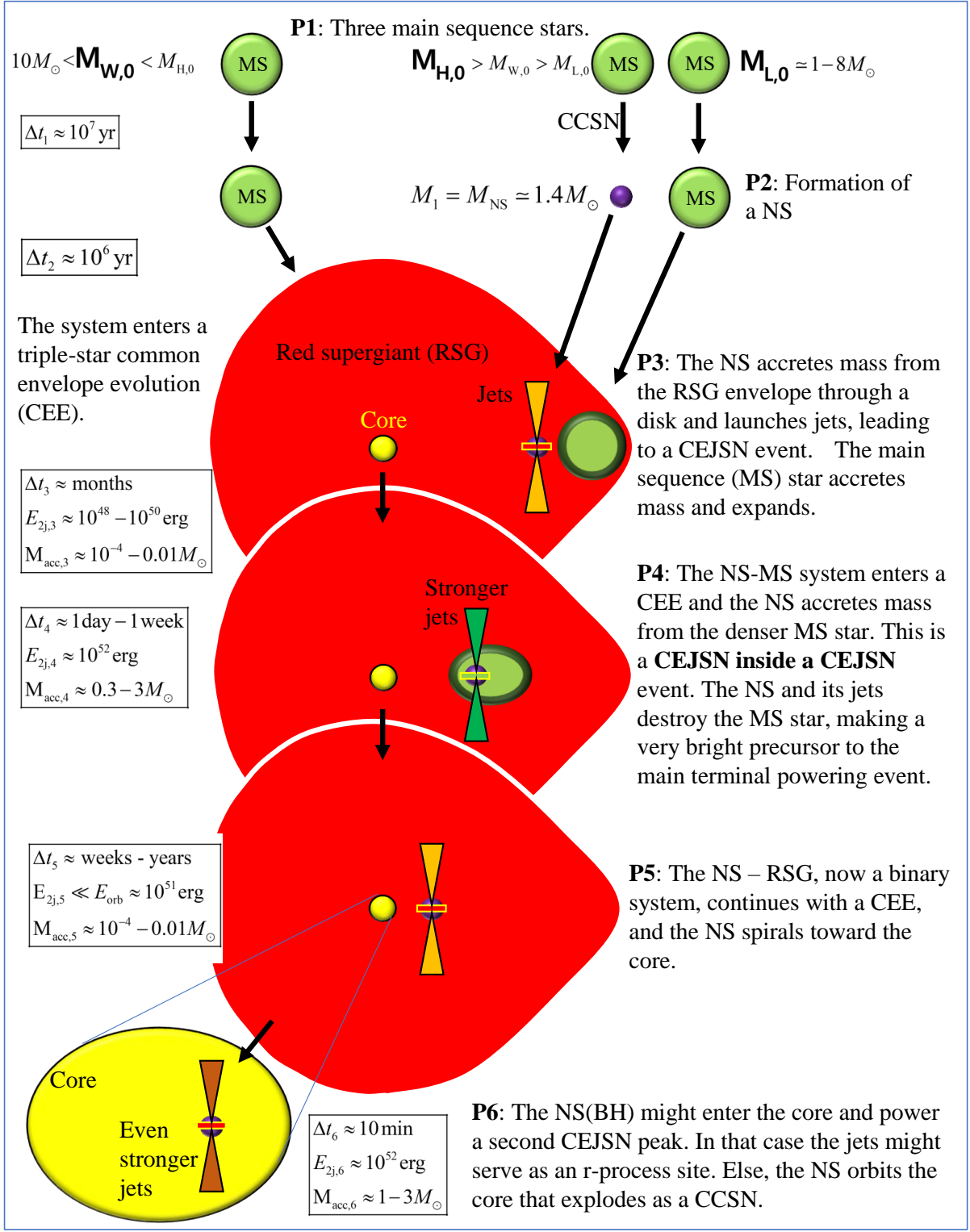}
% \includegraphics[]{CCSNIaFigure.pdf}\\
% \vskip -6.00 cm
\caption{A schematic diagram of the proposed double CEJSN scenario. 
The boxes on the left list the duration of each phase  and for the last four phases the energy that the two jets that the NS launches carry and the mass that the NS accretes, respectively. In the fourth phase the NS might collapse to a BH, and much more likely so in the last phase. P1-P6 stand for the six phases.   
Abbreviation: BH: black hole; CCSN: core-collapse supernova; CEE: common envelope evolution; MS: main sequence; NS: neutron star; RSG: Red supergiant. }
 \label{fig:SchematicScenario}
 \end{figure*}

In this section I discuss the different evolutionary phases of the double CEJSN event that I propose, which I schematic present in Fig. \ref{fig:SchematicScenario}. 
In the double CEJSN scenario there are four types of CEE. (1) The tight binary system enters the RSG envelope. (2) The NS enters the envelope of its MS companion, and this systems continues to spiral-in inside the RSG envelope. The system experiences a CEE within a CEE. (3) The NS/BH remnant spirals-in inside the RSG envelope.  (4) The NS/BH remnant enters the core of the RSG for the final CEE (this occurs if the NS/BH remnant does not remove the entire envelope before it reaches the core). 
%FFFFFFFFFFFFFFFFFFFFFFFFFFFFFFFFFFFFFFFFF
  %[ht]
 % %FFFFFFFFFFFFFFFFFFFFFFFFFFFFFFFFFFFFFF
   
\textit{Phase 1: The initial triple-star system.} The initial state is of a triple-star system of three main sequence (MS) stars, at least two of them massive (CCSN progenitors). The most massive of them, of mass $M_{\rm H,0}$, is in the tight (inner) binary system of this triple system, and the second most massive star, of mass $M_{\rm W,0}$, is on a wide orbit with the tight binary system. The third star of initial mass $M_{\rm L,0}$ might be less or more massive that $8 M_\odot$, but for most cases I study here where the most massive star forms a NS, it is less massive than $8M_\odot$. I estimate the requirement on the initial semi-major axes to be $a_{\rm tb} \la 0.5 \AU$ for the tight binary system, and $a_{\rm wide} \approx 3-10 AU$ for the wide one (larger eccentricities allow higher values of $a_{\rm wide}$). This system seems to be stable (e.g., \citealt{AarsethMardling2001}). This phase is marked by  P1 in Fig. \ref{fig:SchematicScenario}. 
  I estimate these values from the following considerations. I require the wide semi-major axis to be sufficiently small for the wide star to engulf the tight binary system during its RSG phase. I estimate this constraint to be $a_{\rm wide} \la 10 AU$. I also require the tight binary system to form a NS-MS binary with a short orbital period so that it later enters a NS-MS CEE. For that, the tight binary system cannot be too wide such that it avoids both a CEE and a strong tidal interaction. I estimate this constraint to be $a_{\rm tb} \la 0.5 \AU$. For the system to be stable, the wide star cannot be too close, namely, 
$a_{\rm wide} \ga {\rm several} \times a_{\rm tb}$ (depending on the eccentricities and on the mutual inclination of the two orbital planes).  This sets the lower limit on the side semi-major axis.    

\textit{Phase 2: The formation of the NS (BH).} The post-MS evolution brings the most massive star, of initial mass $M_{\rm H,0}  \ga 10 M_\odot$, to explode as a CCSN  after an evolutionary time of $\Delta t_1 \simeq 10^7 \yr$,   and to form either a NS or a BH. Before explosion it transfers mass to the two other stars so that it explodes as a stripped-envelope CCSN (Type Ib or Ic) and ejects little mass, so that the tight binary system stays bound. For example, the lowest-mass star enters the envelope of the most massive star as it becomes a RSG, accretes some of the envelope mass,  $\approx 0.1M_{\rm H,0} - 0.5 M_{\rm H,0}$,   and ejects the rest of the envelope during the CEE. The wide-orbit star might accrete a large fraction of this ejected mass,  crudely up to $10\% - 50\%$ (or even more) of the ejected mass. Namely, the wide star might crudely accrete $\approx 1 M_\odot - 0.4 M_{\rm H,0}$.   
After this phase the mass of the wide-orbit star is $M_{\rm W} > 10 M_\odot$.
As we see below  (section \ref{subsec:Powering}),   more favourable to the present scenario would be $M_{\rm W} \ga 20 M_\odot$ as a more massive wider companion is more likely to bring the NS into its core as it become a RSG star. At the end of the first CCSN the triple-star system is of a tight binary system of a MS star and a NS/BH, and a wide-orbit massive star (second line in Fig. \ref{fig:SchematicScenario} marked by P2).    
In the present study I concentrate on a NS, but a similar scenario with more massive stars might hold for the formation of a BH in this first CCSN.
The tight binary system is similar to the progenitors of low and intermediate X-ray binaries (e.g., \citealt{Podsiadlowskietal2002, TaurisvandenHeuvel2006}), but there is no need for a mass transfer from the MS star to the NS star.
 
 To have the double CEJSN evolution this phase must end with the wide companion more massive than the MS star of the tight binary system, $M_{\rm W} > M_{\rm L}$, so that the wide companion becomes a giant next. For this condition to hold, the initial mass of the wide companion should be substantially larger than that of the lowest star mass star because the lowest mass star might accrete more mass from the most massive star as it evolves first to form the NS. In some case where this phase ends with $M_{\rm W} < M_{\rm L}$, the double CEJSN will not take place.   

\textit{ Phase 3: The beginning of the CEJSN event with a triple-CEE.} The next phase,  at about $\Delta t_2 \simeq 10^6 \yr$ to ${\rm few} \times 10^6 \yr$ after the NS/BH formation,    starts when the wide-orbit star expands to become a RSG star and engulfs the tight binary system (third line in Fig. \ref{fig:SchematicScenario} marked by P3). At early phases in the double-CEJSN event the NS accretes mass from the RSG envelope and launches jets. The powering is as in regular CEJSN (e.g., \citealt{SokeretalGG2019}). If the orbital plane of the tight binary is inclined to the triple-orbital plane the jets are inclined as well, with consequences on the outflow morphology that forms a CSM that lacks any symmetry \citep{Schreieretal2019}. The accretion rate is also different. The more massive tight binary mass, several$\times M_\odot$ compared with $M_{\rm NS} \simeq 1.4 M_\odot$ in the regular CEJSN, attracts more mass to the binary vicinity  (for the simulations of the accretion process by a binary star from an ambient medium see, e.g., \citealt{Comerfordetal2019}).   However, the more massive MS star might accrete most of this mass. 

 Let me elaborate on this. 
\cite{deVriesetal2014} find that when a wide tertiary star overflows its Roche lobe and transfers mass to a tight binary system, the tight binary system ejects most of the mass, and accretes only about fifth of the mass that the tertiary star transfers. The process that I study here involves a NS-MS tight binary that accretes mass inside the envelope rather than a Roche lobe overflow. It is not clear what the dynamical effect of the binary system is on the accretion rate. In any case, I expect that the main reduction in the mass accretion rate results from the removal of envelope mass from the tight binary vicinity by the jets it launches. On the other hand, to have the jets some non-negligible accretion must take place (a negative feedback cycle).  
Future three-dimensional hydrodynamical simulations of this triple-star CEE will teach us on the accretion rate of each of the stars in the tight binary system.  
   
\textit{ Phase 4: The CEJSN inside a CEJSN phase.} When a tight binary system enters a CEE with a giant star there are several classes of outcomes (\citealt{SabachSoker2015, ComerfordIzzard2020, GlanzPerets2021}). The tight binary might survive as it spirals-in, the two stars of the tight binary system might merge inside the envelope (e.g.,  \citealt{Hilleletal2017}), the two stars might enter the core, or the tight binary system might break up. The last class leads to several sub-classes according to the fate of the two stars of the tight binary system, e.g., whether one of the stars leaves the system or whether one or two of the stars merge with the core. 
Relevant to the present study is the following merger evolution (fourth line in Fig. \ref{fig:SchematicScenario} marked by P4). Due to the expansion of the MS star as it accretes mass (because it has a radiative envelope) and/or the gravitational friction inside the RSG envelope, the NS and the MS star merge to start a CEE. Now the NS accretes mass from the MS star, namely, from a much denser environment than that of the RSG envelope (section \ref{sec:Powering}). This results in the launching of very energetic jets that lead to a peak in the light curve. This process ends with the formation of an accretion disk around the NS from the destroyed MS star. The lifetime of the accretion disk is about the viscosity time of the disk which itself is about tens times the dynamical time of the MS star. This entire phase lasts for a time period of (very crudely) $\Delta t_4 \approx 1~{\rm day} - 1~{\rm week}$. Because the total energy of this phase is similar to that of a CEJSN where a NS/BH enters the core of a RSG (section \ref{sec:Powering}), I also term this phase a CEJSN. This merger is different than that of a NS/BH merger with a star outside a RSG envelope, as the massive envelope and CSM ensure a long-duration peak in the light curve that is similar to a CCSN. A NS/BH merger with a MS star in an optically thin medium will lead to a bluer and shorter event.  

\textit{Phase 5: Binary-remnant evolution.} Now the system is of a NS (or a BH) and a RSG in a CEE, as in regular CEJSNe (fifth line in Fig. \ref{fig:SchematicScenario} marked by P5). The NS/BH might eject the entire envelope and leave behind a stable core-NS/BH binary system, where later the core explodes as a CCSN and the system ends as two NSs (or a NS with a BH), bound or unbound. 

\textit{Phase 6: NS-core merger.} The other possibility to end the CEE is that the NS/BH enters the core and accretes mass at a very high rate, leading to a very bright CEJSN peak. In the later case the jets might serve as a r-process site \citep{Papishetal2015, GrichenerSoker2019a, GrichenerSoker2019b}. 

Observations will classify a double-CEJSN event as peculiar super-energetic SN. It is peculiar due to the very long duration, months to few years, very massive ejecta, up to tens of $M_\odot$, very high energy of  $\approx 10^{52} - 10^{53} \erg$, and likely (but not necessarily) with two main peaks in the light curve.

% ==========================================================
\section{Timescales and powering} 
\label{sec:Powering}
% ===========================================================

\begin{figure*}
\includegraphics[trim= 1.0cm 15.5cm 1.0cm 2.5cm,clip=true,width=1.00\textwidth]{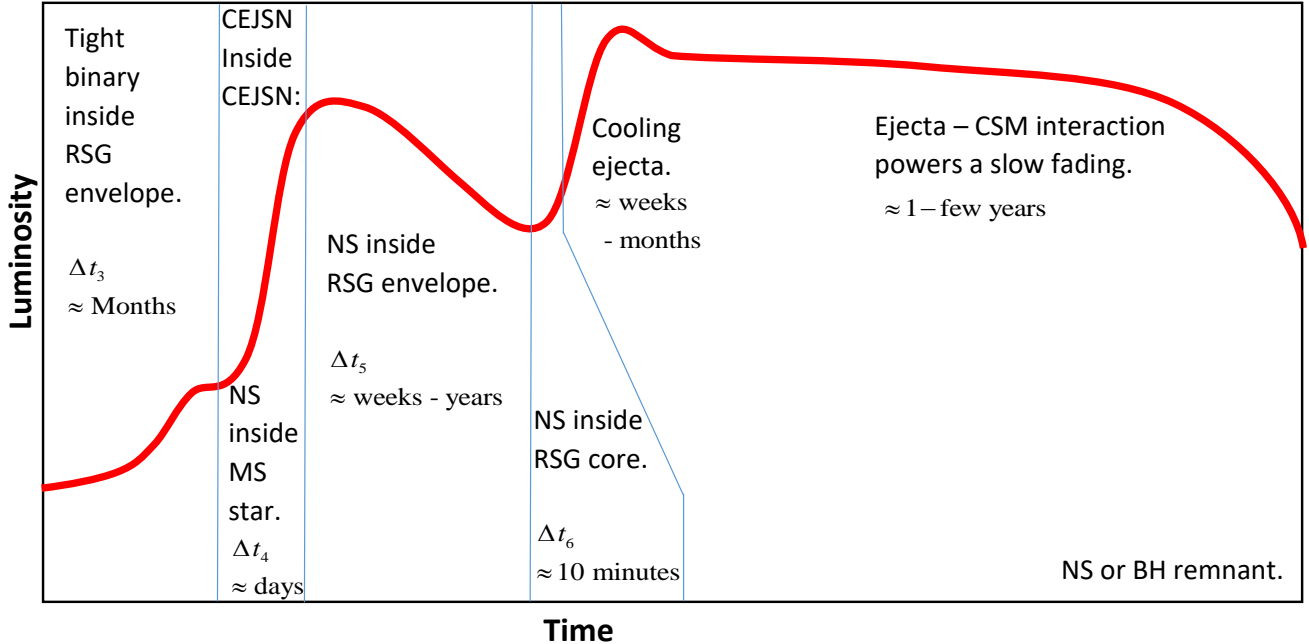}
\caption{  A schematic drawing (not to scaled) of the expected light curve of the double CEJSN event. The first four phases are the last four phases of Fig. \ref{fig:SchematicScenario}. If the NS does not enter the core, the core explodes as a CCSN. The light curve beyond the phase of a NS inside the RSG envelope depends in that case (not shown) on the time from the ejection of the envelope to the explosion, i.e., the properties of the CSM at the explosion time. The total energy, kinetic and radiation, that the jets carry in their interaction with the envelope and CSM that leads to each of the two large peaks in the light curve is $\approx 10^{52} \erg$.      
}
\label{fig:LC}
\end{figure*}

In this section I emphasise the powering role of jets. In addition there is the gravitational energy that the system releases in the CEE processes. A large fraction of the orbital energy goes to unbind the envelopes, e.g., the recent 3D numerical simulation of a NS-RSG CEE by \cite{LawSmithetal2021} and \cite{Schroderetal2020}.
As for the energy that accretion liberates via jets, which also depends on the response of the NS to accretion \citep{Holgadoetal2021}, it can be about comparable to or much larger than the orbital energy that the in-spiralling system release.  
 I schematically (not to scale) present the expected light curve in Fig. \ref{fig:LC}.  
% FFFFFFFFFFFFFFFFFFFFFFFFFFFFFFFFFFFFFFFFFFFFFFFFFFFFFFFFFFFFFF
% FFFFFFFFFFFFFFFFFFFFFFFFFFFFFFFFFFFFFFFFFFFFFFFFFFFFFFFFFFFFFF

The CEE proceeds in several phases (e.g., \citealt{IvanovaNandez2016} for a recent discussion). Relevant to the present study are the plunge-in rapid phase and the much slower self-regulating spiral-in phase. 
In the plunge-in phase the compact star reaches deep into the envelope of the large star during about the Keplerian orbital time on the surface of the giant star (e.g., \citealt{Termanetal1995, TaamRicker2010, Passyetal2012, RickerTaam2012, IvanovaNandez2016, Ohlmannetal2016, Shiberetal2019, Chamandyetal2020, Krameretal2020, Reichardtetal2020, Sandetal2020, GlanzPerets2021}). The self-regulating phase lasts for a longer time. The same behavior of CEE in binary systems applies for triple star systems, when observing the distance between the center of mass of the tight binary system and the core of the RSG \citep{GlanzPerets2021}. 

In what follows, the subscripts correspond to the   phases P3-P6 in Fig. \ref{fig:SchematicScenario} that I also described in section \ref{sec:Scenario}.   For example, phase 4 is the CEE of the NS inside the MS star.  

% =========================
\subsection{Tight binary system inside the RSG envelope} 
\label{subsec:Powering}
% =========================

During the plunge-in time the NS accretes mass and launches jets (P3 in Fig. \ref{fig:SchematicScenario}). The accretion rate of a single NS in a RSG envelope is $\dot M_{\rm acc} \approx 0.1 - 0.2 \dot M_{\rm BHL}$, where $\dot M_{\rm BHL}$ is the Bondi-Hoyle-Lyttleton mass accretion rate (e.g., \citealt{LopezCamaraetal2020MN}). Due to negative feedback from the jets (as they remove mass from the envelope), the mass accretion rate in the loosely bound envelope is expected to be smaller even. The total energy that the jet carry might be several times the binding energy of the envelope (\citealt{SokeretalGG2019}; a large fraction of the energy of the jets goes to accelerate the ejected envelope gas to velocities larger than the escape velocity). For a RSG, the binding energy of the envelope above radius $r$ depends on the RSG radius $R_{\rm RSG}$ and its mass. Typical values for the binding energy of the envelope residing above $r \ga 10 R_\odot$ are $E_{\rm bind} (10 R_\odot) \approx 10^{47} - 10^{49} \erg$, where the lower values are for RSG masses of $M_{\rm RSG} \approx 10 M_\odot$ and radii of $R_{\rm RSG} \simeq 10^3 R_\odot$ and the larger values are for RSG of $M_{\rm RSG} \approx 30 M_\odot$ and $R_{\rm RSG} \simeq 200 R_\odot$ (e.g., \citealt{Lohevetal2019}).
 
If the two jets together carry a fraction $\eta  \simeq 0.1$ of the accreted mass at a velocity of $v_j \simeq 10^5 \km \s^{-1}$, and these jets carry a total energy of $q$ times the envelope binding energy (see \citealt{SokeretalGG2019}), the total mass that the NS accretes in this phase is 
\begin{eqnarray}
\begin{aligned}
M_{\rm acc,3} \approx  10^{-3} 
\left( \frac{q}{10} \right)
\left( \frac{\eta}{0.1} \right)^{-1}
\left( \frac{E_{\rm bind}}{10^{48} \erg} \right)  
M_\odot  .
%\left( \frac{}{} \right)
\end{aligned}
\label{eq:Macc3}
\end{eqnarray}

I assume that during the plunge-in phase or shortly after it the tight binary suffers a strong enough perturbation, by the expansion of the MS star as it accretes mass and by gravitational friction, that they merge. Namely, the NS-MS merger occurs at a time $\Delta t_3 \approx \Delta t_{\rm plunge}$. This is about the Keplerian orbital time on the surface of the RSG of radius $R_{\rm RSG}$ 
\begin{eqnarray}
\begin{aligned}
\Delta t_3 \approx  0.1 - 0.2 
\left( \frac{R_{\rm RSG}}{200 R_\odot} \right)^{3/2} 
\left( \frac{M_{\rm total}}{30 M_\odot} \right)^{-1/2}
\yr ,
%\left( \frac{}{} \right)
\end{aligned}
\label{eq:DeltaT3}
\end{eqnarray}
where $M_{\rm total}$ is the total mass of the triple system at the onset of the CEE. 

From equations (\ref{eq:Macc3}) and (\ref{eq:DeltaT3}) the typical accretion rate onto the NS in this phase is $\dot M_{\rm acc,3} \approx 0.01  M_\odot \yr^{-1}$. This is high enough to allow neutrino cooling \citep{HouckChevalier1991}. 
In the relevant zones of $r \approx 10-100 R_\odot$
the envelope density is $\approx 10^{-6} - 10^{-4} \g \cm^{-3}$ and the Bondi-Hoyle-Lyttleton mass accretion rate of a single NS is very crudely $\dot M_{\rm BHL} \approx 0.01-0.1 M_\odot \yr^{-1}$.  
\cite{LopezCamaraetal2019} and \cite{LopezCamaraetal2020MN} find the NS to accrete at a rate of $\dot M_{\rm acc} \approx 0.1 - 0.2 \dot M_{\rm BHL}$, and so the NS can indeed accrete at the required rate (here there is the complication of the presence of the MS close companion). 

Overall, the phase of tight binary inside the RSG in the double CEJSN scenario lasts for months to few years, and the energy that the two jets carry is  
\begin{equation}
E_{\rm 2j,3} \approx 10^{48} -10^{50} \erg, 
\label{eq:E3}
\end{equation} 
with an expected large variations between different systems.  During this phase the NS accretes too little mass to become a BH. 
 Note that, as mentioned above, the binding energy of the envelope residing above $\simeq 10 R_\odot$ is $\approx 10^{47} \erg$ for lower mass RSG progenitors of CCSNe, and up to $\approx 10^{50} \erg$ for the envelope residing above a radius of $\simeq 0.1R_\odot$ of more massive CCSN progenitors (e.g., \citealt{Lohevetal2019}).   Therefore, the orbital energy that the triple system releases during this phase might be up to about few times the envelope binding energy (e.g., as in the NS-RSG simulation of \citealt{Termanetal1995}), and so it is typically smaller than the energy the jets carry. In some case it will be smaller but not by much, and in some other cases it might even be comparable and somewhat larger than the energy the jets carry. 

% =========================
\subsection{A NS-MS merger inside the RSG envelope} 
\label{subsec:Merger}
% =========================

This phase is a new ingredient I introduce into the family of the CEJSN scenario. 
Unlike giant stars, the MS star does not have the structure of a tenuous envelope and a very dense core. The CEE brings the NS close to the center of the MS, and due to its huge gravity the NS tidally destroys the MS star.  For a NS inside a MS star this takes place when the stellar mass inner to the NS orbit is about equal to or less than the NS mass (because then the tidal disruption radius becomes larger than the orbit), as in a micro-tidal disruption event (micro-TDE) where a NS/BH disrupts a star \citep{Peretsetal2016}.   Because of the large angular momentum of the binary system the destroyed MS gas forms an accretion disk around the NS. 
The first part of the spiralling-in lasts for about the Keplerian time on the surface of the MS star $t_{\rm{Kep}}^{\rm MS}$, about few hours. The accretion of the rest of the destroyed MS gas lasts for about the viscous timescale of the accretion disk. 
In the $\alpha$ disk model where the viscosity is $\nu = \alpha_{\rm d} C_s H$, where $C_s$ is the sound speed and $H(R)$ the disk height at radius $R$ from the NS, the viscous timescale is given by
\begin{equation}
%\begin{split}
t_{\rm{visc}}^{\rm MS} =\frac{R^2}{\nu} = \frac{R^2}{\alpha C_s H} \simeq
t_{\rm{Kep}}^{\rm MS} \left(\frac{R}{H}\right)^{2} \frac{1}{2 \pi \alpha_{\rm d}}.   % \\
% =30 \left(\frac{R}{10 H}\right)^{2}
% \left(\frac{R_2}{20R_{\odot}}\right)^{3/2} \left(\frac{M_2}{30
% M_{\odot}}\right)^{-1/2} \frac{1}{\alpha}\days , 
\label{eq:Tvisc}
%\end{split}
\end{equation}
The large uncertainty here is in the ratio $R/H$. In a thin accretion disk the typical value is $(R/H) \simeq 10$. However, here the destroyed MS starts as a sphere, and it takes time for the disk to collapse to the equatorial plane. Another unknown parameter is the viscosity coefficient $\alpha_{\rm d}$.  
It is likely that over several dynamical timescales the outer edge of the disk expands, increasing the dynamical time of the outer parts of the disk, and the disk becomes thinner, increasing the value of $R/H(R)$. Both these effects increase the viscous timescale. Overall, this accretion phase decays over a timescale of tens to about a hundred times the Keplerian orbital time of the MS star, or about a day to a week
\begin{equation}
%\begin{split}
\Delta t_4 \approx t_{\rm{visc}}^{\rm MS} \approx 1 
\left( \frac{t_{\rm{Kep}}^{\rm MS}}{ 2 {\rm h}} \right) \left(\frac{R}{3H(R)} \right)^{2} \frac{1}{2 \pi \alpha_{\rm d}}   {\rm day}.   % \\
% =30 \left(\frac{R}{10 H}\right)^{2}
% \left(\frac{R_2}{20R_{\odot}}\right)^{3/2} \left(\frac{M_2}{30
% M_{\odot}}\right)^{-1/2} \frac{1}{\alpha}\days , 
\label{eq:Deltat4}
%\end{split}
\end{equation}

The fraction of the MS mass that the NS accretes depends on how efficient the polar jets are in removing the mass. The NS is likely to accrete up to about half the mass, or $M_{\rm acc,4} \approx 0.3- 3 M_\odot$. This might bring the NS to collapse to a BH. Over all, the accretion rate and the total energy that the two jets carry are 
\begin{eqnarray}
\begin{aligned}
& \dot M _{\rm acc,4} \approx 10 - 1000 M_\odot \yr^{-1} .  
\\ &
E_{\rm 2j, 4}  \approx 10^{52} \left(\frac{\eta}{0.1} \right) 
\left( \frac {M_{\rm acc}}{1 M_\odot} \right) \erg,  
\end{aligned}
\label{eq:Lphase4}
\end{eqnarray}
respectively. 

The jets will eject most of the RSG envelope outer to the orbit of the NS or the  BH remnant, but some equatorial gas might stay bound. The outflow will be very fast, but it will collide with the CSM that the system ejected in the previous phase. The collision of the ejecta with the CSM will transfer a large fraction of the kinetic energy to radiation (e.g., \citealt{Schroderetal2020}), forming a super-luminous event, even if the NS accretes only $0.1 M_\odot$. The peak in the light curve will last for months to a year due to the long photon diffusion time in the massive CSM (Fig. \ref{fig:LC}).     
 
Some of the MS gas might end as bound gas, i.e., it becomes part of the common envelope (RSG envelope).  
 
If the NS-MS merger event ejects most of the envelope, such that the NS or BH remnant does not spiral-in to the core, the event ends here, until much later  (more than hundreds of years) the core explodes as a CCSN. 

% =========================
\subsection{CEE after the destruction the MS star} 
\label{subsec:PostMerger}
% =========================
 
Now the system proceeds as a CEJSN of a binary star, a NS/BH and an RSG. For example, if the remnant is a black hole the relativistic jets it launches can be a strong source of  $\approx 10^{15} \eV$ neutrinos \citep{GrichenerSoker2021}. 

In the double CEE event the NS-MS merger process and its jets do not remove the entire envelope, and the RSG envelope that is still bound is massive enough to cause the NS/BH remnant to spiral-in to the core. As said, part of the MS mass might mix with the original RSG envelope. The NS/BH remnant starts this phase very close to the core, orbital separation much smaller than the initial radius of the RSG star. However, the envelope might be synchronised with the orbital motion and the spiralling-in phase might be many times the Keplerian orbital period.   It is very hard to estimate the duration of this phase, the accretion rate, and the total power. But due to the lower envelope mass, the jets that the NS/BH launch in this phase do not add much to the energy of the event. The main energy source of this phase is the orbital energy $E_{\rm orb}$ of the core-NS/BH binary systems that spirals-in to merge. 

 The duration of this phase and the energy of the jets are highly uncertain. To give some plausible values I assume the following. (1) I assume that when the mass of the reminding envelope is large so is the dynamical friction and the spiralling-in time to the core, i.e., of the order of the orbital period where the previous phase (NS-MS CEE) took place $a_4$, about two weeks for $a_4 \simeq 50 R_\odot$. If the reminding envelope is of low mass, the evolution might be much longer and take years. For example, if the envelope shrinks to be less that the orbital separation $a_4$, then the evolution is due to tidal interaction that can take tens of orbital periods. (2) I assume that the jets operate in a negative feedback cycle. This implies that if the jets are too strong they remove large parts of the envelope and accretion decreases. In addition, in the spiralling-in towards the core the NS releases a large amount of energy that by itself can unbind most of the envelope and leads to the decreasing of the accretion rate. I assume therefore, that at this phase the released orbital energy plays a larger role that the jets, i.e., releases more energy that the jets carry.   
Very crudely I summarise the duration of this phase and the jets power as
\begin{eqnarray}
\begin{aligned}
& \Delta t_5 \approx {\rm weeks - years} 
\\ &
E_{\rm 2j, 5} \ll E_{\rm 2j, 4} \qquad {\rm and} 
\\ &
E_{\rm 2j, 5} \ll E_{\rm orb} \simeq 7 \times 10^{50}
\left( \frac{M_{\rm core}}{5 M_\odot} \right)
\left( \frac{M_{\rm NS}}{1.4 M_\odot} \right)
\\ &  \qquad \qquad \qquad 
\times 
\left( \frac{a_5}{0.02 R_\odot} \right)^{-1}
\erg , 
\end{aligned}
\label{eq:Lphase5}
\end{eqnarray}
where $M_{\rm core}$ and $a_5$ are the core mass and final orbital separation at this phase before the final spiralling-in into the core starts.

% =========================
\subsection{NS/BH inside the core} 
\label{subsec:InsideCore}
% =========================
 
When the NS/BH enters the core we are in the regular process of a CEJSN. It is hard to estimate the accretion rate of this process due to the complicated gravitational interaction and the role of jets. Unlike the case of NS star merging with a WD, here the core is more massive than the NS. For that, the Bondi-Hoyle-Lyttleton mass accretion rate might be a crude estimate at the early times of the NS-core merger (but not for a BH that is more massive). 
\cite{GrichenerSoker2019a} find this rate to be $\dot M_{\rm BHL} \approx 0.01-0.1 M_\odot \s^{-1}$. 
A BH will tidally destroy the core in at early times, a process similar to a  micro-TDE event where a NS/BH merges with a star \citep{Peretsetal2016}.  
I estimate the mass accretion as I did in equation (\ref{eq:Deltat4}), but now the Keplerian orbital time is that of a core, e.g., $t_{\rm{Kep}}^{\rm core} \simeq 100 \s$. According to \cite{GrichenerSoker2019a} the highest value of $\dot M_{\rm BHL}$ is when the NS is at $a \simeq 0.01-0.03R_\odot$ from the center of the core and the inner core mass  is $\approx 1M_\odot$. I take this last phase when the inner mass is about that value, and so the Keplerian time of this inner core is (for a NS of mass $1.4 M_\odot$) $t_{\rm{Kep}}^{\rm in} \simeq 5-30 \s$.   
The viscous timescale, which I also take to be the duration of the highest accretion rate phase from the core, is then 
\begin{equation}
\Delta t_6 \approx t_{\rm{visc}} \simeq 100 
\left( \frac{t_{\rm{Kep}}^{\rm in}}{ 10 \s } \right) \left(\frac{R}{3H(R)} \right)^{2} \frac{1}{2 \pi \alpha_{\rm d}}  \s . 
\label{eq:Deltat6}
\end{equation}

The NS/BH accretes $\approx 1 M_\odot$ \citep{GrichenerSoker2019a}, which is most of the core (as the jets eject only the polar regions). 
Before that phase, for about hundreds to few thousands seconds the NS/BH might accrete mass from the outer parts of the envelope. As well, it might accrete fallback gas at later times. However, the highest accretion rate lasts for about 100 seconds and the accretion rate and total energy that the two jets carry are 
\begin{eqnarray}
\begin{aligned}
& \dot M _{\rm acc,6} \approx 0.01 M_\odot \s^{-1} ,  
\\ &
E_{\rm 2j, 6}  \approx 10^{52} \left(\frac{\eta}{0.1} \right) 
\left( \frac {M_{\rm acc}}{1 M_\odot} \right) \erg,  
\end{aligned}
\label{eq:Lphase6}
\end{eqnarray}
respectively.  
  
The large energy of $E_{\rm 2j, 6}  \approx 10^{52} \erg$ forms a second energetic peak in the light curve of the long-lasting event. Due to the high opacity of the newly ejected core gas and the previously ejected MS gas and RSG envelope gas, the optical depth of the entire system is large, and the second peak might last for several years. If the jets break through the polar directions, on the other hand, then the second peak might be much shorter than a year, similarly to the polar-CEJSN scenario that \cite{SokeretalGG2019} proposed for the fast-rising blue optical transient AT2018cow.

The accretion rate $\dot M _{\rm acc,6}$ is the average value over about 100 seconds. It is about 0.1 times the highest value of $\dot M_{\rm BHL}$ that \cite{GrichenerSoker2019a} find, which is expected as the largest values is only in a specific radius inside the core. I find that the two estimates of the mass accretion rate inside the core are compatible. Moreover, these values are compatible with the results of \cite{ZhangFryer2001} when I scale to a higher core-density than what they use (as they take a helium core and not a CO core as I assume here).
The average accretion rate of $\approx 0.01 M_\odot$ is above the threshold of $0.002 M_\odot \s^{-1}$ for r-process nucleosynthesis according to \cite{Siegeletal2019}. This process might be a source of r-process nucleosynthesis \citep{GrichenerSoker2019a}.  

The above discussion was for a RSG core after core helium burning. In case of a helium core, the density is lower and the mass accretion rate is lower (e.g., \citealt{Siegeletal2019}), and it is below the threshold for the r-process nucleosynthesis. The total energy that the jet carry, thought, might be the same. 

% ==========================================================
\section{Event rate} 
\label{sec:rate}
% ===========================================================

\cite{Schroderetal2020} performed population synthesis calculations and found that for RSG stars of mass $\ga 10 M_\odot$ the frequency of NS/BH entering a RSG envelope is about $1.5 \times 10^{-4} M^{-1}_\odot$, i.e., rate per solar mass of stars formed. This rate is $2.6\%$ of the CCSN rate  (\citealt{Chevalier2012} estimate this rate to be about $1\%$).  For the CCSN rate in the local Universe (e.g., \citealt{Strolgeretal2015}) this rate of NS/BH entring RSG envelope is $ \simeq 2 \times 10^{-6} \yr^{-1} \Mpc^{-1}$.    From these, \cite{Schroderetal2020} find that in about $22 \%$ the NS/BH enters the RSG core, i.e., a CEJSN event. In the present nomenclature the rest, about $78 \%$ are CEJSN impostors that form NS/BH binaries that later might merge as a gravitational waves source. \cite{Schroderetal2020} find that  in about half of all CEJSNe the compact object is a NS and in the other half it is a BH. In total, the CEJSN rate they find is $0.6\%$ of the rate of CCSNe. \cite{GrichenerSoker2019a} use the r-process CEJSN scenario to argue that the rate of NS entering a CO core of a RSG should be about $0.1 \%$ of the CCSN rate. Other cases where a NS enters the RSG core before helium burning do not lead to r-process. 
These different estimates are compatible. 
I take the CEJSN rate to be $f_{\rm CEJ} \simeq 0.006$ times the CCSN rate, and CEJSN cases with post-helium burning RSG core to be about $\simeq 0.3-0.5$ of those cases, namely a ratio of $f_{\rm CEJ,He} \simeq 0.002$ to CCSN rate. 

Let me now estimate the number of cases where instead of being a single NS/BH, the NS/BH is in a tight binary system with a MS star. I assume that this fraction is about equal to the fraction of binary systems where a NS/BH has a MS companion. 
 I base this assumption on that the evolution of the most massive star in the triple did not eject its close MS companion, and so most likely it did not eject the wider star. As well, I assume that the fraction of NS-MS close binary systems does not depend on the presence of a wide companion.   
\cite{Pfahletal2003} estimate the birthrate of low- and intermediate (MS mass of $<4M_\odot$) X-ray binaries to be $\approx 10^{-4} - 10^{-2}$ of the CCSN birthrate. 
\cite{ShaoLi2015} estimate the birth rate of all X-ray binaries to be a fraction of $\approx 0.01$ of CCSN birthrate, most of them with a MS star of $M> 5 M_\odot$. 
Overall, I take that, very crudely, $f_{\rm tb} \approx 0.003$ of all NS form the tight binary systems that I require in this study  (the range can be $f_{\rm tb} \approx 10^{-3}-0.01$).   The fraction of NS in low-mass X-ray binaries can be as low as $f_{\rm tb}({\rm low}) \approx 0.001$, but the fraction of high mass X-ray binaries can be two orders of magnitude higher $f_{\rm tb}({\rm high}) \approx 0.1$ \citep{TaurisvandenHeuvel2021}. 
\cite{ShaoLi2020} claim that X-ray binaries with a BH companion favour low mass BHs. The formation rate is less certain, and I take it to be the same as that of X-ray binaries with a NS companion. 

My crude estimated ratio of double CEJSN event rate with a NS or BH companion to CCSNe rate is 
\begin{equation}
f \approx f_{\rm tb} f_{\rm CEJ} \approx 6 \times 10^{-6} - 6 \times 10^{-5}. 
\label{eq:fDCEJ}
\end{equation}
 If one to take one value, I crudely estimate $f \approx 2 \times 10^{-5}$.  
About half of these are with a NS star, what I term here double CEJSN, i.e., $f_{\rm DCEJ} \approx 10^{-5}$.
Half of these cases are with a BH, what I term a micro-TDE CEJSN, i.e., $f_{\rm BH,CEJ} \approx  10^{-5}$. The rate of double CEJSN events with a NS companion and a CO core of the RSG is 
$f_{\rm DCEJ}{\rm (CO)} \approx 3 \times 10^{-6}$.
    
With new and future transient surveys,  e.g., All-Sky Automated Survey for Supernovae (ASAS-SN; \citealt{Kochaneketal2017PASP}), The Zwicky Transient Facility (ZTF; \citealt{Bellmetal2019}), and the Southern Hemisphere Variability Survey (LSQ; \citealt{Baltayetal2013}),    which might detect about $10^4$ CCSNe per year, I very crudely expect the detection of one event per year that is a double CEJSN or a micro-TDE. The reason that the ratio of $\approx 10^{-4}$ to CCSN detection rate is larger than the value given by equation (\ref{eq:fDCEJ}) is that the double CEJSN events are much brighter than typical CCSN events, and therefore the volume coverage is much larger.  

% ==========================================================
\section{Summary} 
\label{sec:Summary}
% ===========================================================

I presented the basic phases,  including timescales, energy, and NS mass accretion,   of the double CEJSN in Fig. \ref{fig:SchematicScenario}. In equations (\ref{eq:Macc3})-(\ref{eq:Lphase6}) I gave my crude estimates of the timescales and jet-powers of the four CEE phases when the NS (or BH) launch jets.
 In proposing this scenario I made the following main assumptions. 
(1) I assumed that when a tight NS-MS binary system enters a RSG envelope the interaction with the RSG envelope can bring the NS-MS binary system to merge. One process that might prevent the merger is if the jets that the NS launches remove too much gas from the NS-MS binary vicinity and the interaction of the NS-MS binary with the envelope is not sufficiently strong to bring the NS and MS star to merge. Namely, the NS-MS binary system performs the grazing envelope evolution rather than the CEE (e.g., \citealt{ShiberSoker2018}).
(2) I assumed that the NS-MS merger process ends with a large fraction of the MS mass forming an accretion disk around the NS, and that the NS accretes a large fraction of the disk mass. Here, again, the early jets might efficiently remove most of the MS stellar mass such that the total accreted mass is very low, i.e., $M_{\rm acc,4} \la 10^{-2} M_\odot$.

 For these uncertainties,   all these estimates require better determinations by future studies, mainly of complicated and highly demanding 3D hydrodynamical simulations. 

In section \ref{sec:rate} I crudely estimated the rate of the double CEJSN events to the rate of CCSNe. About 1 double CEJSN event takes place per $10^5$ CCSNe. Since CEJSNe are much brighter than typical CCSNe, their detection can cover a larger volume and the detection rate, including also the micro-TDE CEJSN events where there is a BH instead of a NS might be $1:10^4$.
This implies that new transient surveys might detect in total one triple-star CEJSN event per year. 

The double CEJSN event is one of several different types of triple-star CEJSN events (cases). I end by calling the attention of future discoverer of peculiar super-energetic SNe to the CEJSN scenario, in binaries and in triples.

% ===================================================
\section*{Acknowledgments}
% ===================================================

I thank Aldana Grichener for helpful discussions and comments. 
 I thank an anonymous referee for detailed comments and helpful suggestions.  
This research was supported by the Amnon Pazy Research Foundation.

%%%%%%%%%%%%%%%%%%%%%%%%%%%
\textbf{Data availability}
The data underlying this article will be shared on reasonable request to the corresponding author.  
%%%%%%%%%%%%%%%%%%%%%%%%%%%

% %%%%%%%%%%%%  References %%%%%%%%%%%%%%%%%%%%%

\label{lastpage}
\end{document}